\documentclass[aps,prb,twocolumn,showpacs]{revtex4-1}
\usepackage{graphicx}
\usepackage{amssymb}
\usepackage{amsmath}
\usepackage{appendix}
\usepackage{comment}

\begin{document}

\title{Thermally-driven spin torques in layered magnetic insulators}

\author{Scott A. Bender and Yaroslav Tserkovnyak}
\affiliation{Department of Physics and Astronomy, University of California, Los Angeles, California 90095, USA}

\begin{abstract}
Thermally-driven spin-transfer torques have recently been reported in electrically insulating ferromagnet$|$normal-metal heterostructures. In this paper, we propose two physically distinct mechanisms for such torques. The first is a local effect: out-of-equilibrium, thermally-activated magnons in the ferromagnet, driven by a spin Seebeck effect, exert a torque on the magnetization via magnon-magnon scattering with coherent dynamics. The second is a nonlocal effect which requires an additional magnetic layer to provide the symmetry breaking necessary to realize a thermal torque. The simplest structure in which to induce a nonlocal thermal torque is a spin valve composed of two insulating magnets separated by a normal metal spacer; there, a thermal flux generates a pure spin current through the spin valve, which results in a torque when the magnetizations of the layers are misaligned.  

\pacs{72.25.Mk, 72.25.Rb, 72.20Pa, 75.76+j}

\end{abstract}

\maketitle

\section{Introduction}

The growing field of spin caloritronics\cite{Bauer:2012to} complements the electrical control of spin current with a new experimental bias: temperature gradient. In contrast to electrical biasing, which couples to the electron charge, transport by the application of a thermal flux is possible for neutral carriers. If, for example, a temperature gradient is applied to a magnetic insulator, a net flow of angular momentum, carried by thermally-activated spin-wave excitations, results.\cite{Uchida:2010ei,*Adachi:2013vc} When integrated into larger structures, magnonically-active elements open the possibility of new effects and devices based on thermally-driven transport. \cite{Chumak:2015fa}

One such effect is that of a thermal spin-transfer torque at a normal-metal$|$insulating ferromagnet interface, which has been recently observed\cite{PadronHernandez:2011da,*Jungfleisch:2013gg,Lu:2012it} via the modulation of ferromagnetic resonance linewidth. Thermally-driven magnetic dynamics were first predicted\cite{Hatami:2007gp,Jia:2011ja,*Heiliger:2014eh,Luc:2014hr} and reported\cite{Choi:2015ed,*Yu:2010kj} for $conducting$ ferromagnetic layers, where the spin-transfer torque can be provided by spin-polarized electric current injected into the magnetic layer by an interfacial spin-dependent Seebeck effect.\cite{Slachter:2010hj,*Hu:2014ck} In contrast, for an insulating ferromagnetic layer, spin-transfer torque can arise only from a thermally-driven pure spin current mediated by ferromagnetic magnons. A general framework, describing the interplay between magnon transport and the ferromagnetic order-parameter dynamics, however, has been lacking. 

In this paper, we provide an account of the physics of thermal magnon-mediated spin-transfer torques arising in normal-metal (N)$|$insulating-ferromagnet (F) heterostructures, building on the formalism developed in our previous works.\cite{Bender:2014hga,Bender:2015jx} In Sec.~\ref{secloc}, we construct a local mechanism, which utilizes SU(2) symmetry breaking of an anisotropic F to couple thermally-activated magnons to the spin-density order parameter. In conjunction with the spin Seebeck effect, this engenders a thermally-driven torque in F. In Sec.~\ref{secnl}, we investigate a nonlocal mechanism, where the SU(2) symmetry is structurally broken. The simplest example of this is an F$|$N$|$F trilayer, in contact with normal-metal leads that serve as reservoirs of angular momentum.  A thermomagnonic flux passing through the ferromagnetic components results in a spin accumulation in the normal-metal spacer, which exerts a torque on the ferromagnetic layers, in close analogy with a traditional electronic spin valve.  For both mechanisms, we obtain the change in magnetic damping in linear response to a temperature gradient and consider magnetic dynamics induced beyond linear response.

\section{Local mechanism} 
\label{secloc}

 For illustrative purposes, we discuss the local mechanism for thermal spin-transfer torque in the simplest possible structure: an N$|$F bilayer. A spin current entering F (assumed to form a single domain) through the N$|$F interface is comprised of two orthogonal, physically distinct components. The first is the spin current collinear with the spin density order parameter unit vector $\mathbf{n}$, with $\mathbf{n}$ taken to be spatially uniform in the thin film limit; physically, this current arises from thermal fluctuations, and on the F side is transported by magnons.  The second current, which is orthogonal to $\mathbf{n}$ and linear in $\mathbf{n}\times \boldsymbol{\mu}'$ (where $\boldsymbol{\mu}'$ is the spin accumulation in N along the interface) and $\dot{\mathbf{n}}$, gives rise to the spin-transfer torque on $\mathbf{n}$.\cite{Slonczewski:1996vc} In the presence of a temperature gradient across the N$|$F interface, a spin current of the first kind flows, which results in the buildup of a thermally induced spin accumulation in N along the interface. Crucially, this spin accumulation is collinear with the order parameter $\mathbf{n}$, and therefore $cannot$ produce a spin-transfer torque on $\mathbf{n}$.  In contrast to the claim of Ref.~\onlinecite{Lu:2012it}, a temperature gradient maintained across an N$|$F interface cannot exert a spin torque on F in the absence of SU(2) symmetry breaking.

\begin{figure}[pt]
\includegraphics[width=\linewidth,clip=]{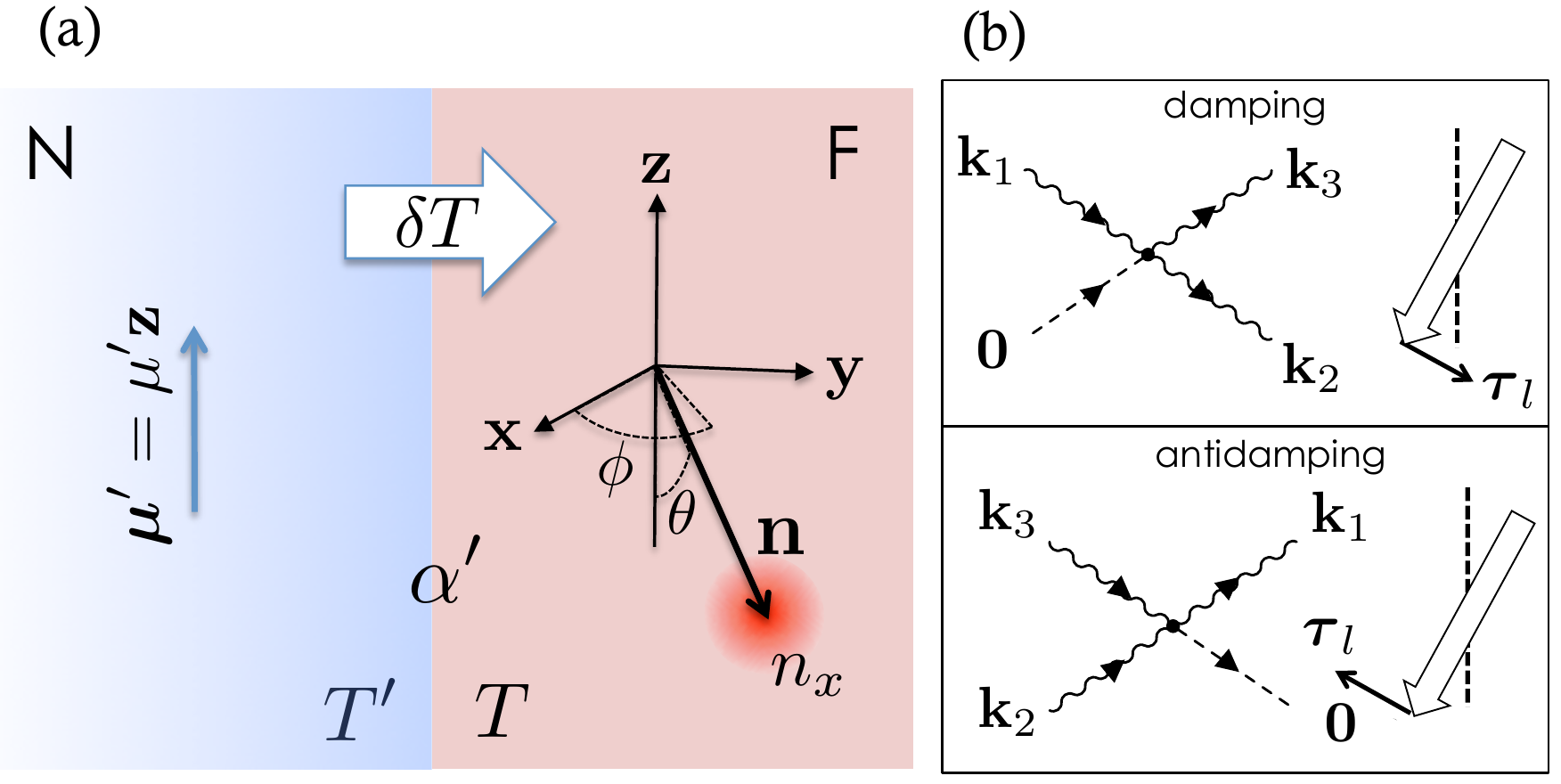}
\caption{a.) Schematic for the N$|$F bilayer. An interfacial temperature drop $\delta T$ drives spin current into F, which is absorbed by the magnons. b.) Magnon scattering processes, opened when the spin density order parameter $\mathbf{n}$ is misaligned with the F broken-symmetry axis $z$.  The annihilation of one finite $\mathbf{k}$ (thermal) magnon and the corresponding creation of two robs $\mathbf{n}$ of $\hbar$ of angular momentum in the $\mathbf{z}$ direction, resulting in a damping torque; the inverse process generates an antidamping torque. Wavy lines represent $\hat{\varphi}$ magnons, while straight, $\Phi$ magnons.}
\label{sch1}
\end{figure}

A thermal spin torque on $\mathbf{n}$ in an N$|$F structure (Fig.~\ref{sch1}) therefore requires SU(2) symmetry breaking by the F layer itself. In the simplest case, which we consider here, this is provided by local uniaxial anisotropy.  The ferromagnetic Hamiltonian is:
\begin{equation}
\hat{\mathcal{H}}=\int d^3 x\left(-\frac{A}{2 s}\hat{\mathbf{s}}\cdot \nabla^2 \hat{\mathbf{s}}+H\hat{s}_z+\frac {K}{2 s} \hat{s}_z^2 \right)\, .
\label{H}
\end{equation}
Here, $A$ is the exchange stiffness, $s$ is the saturation spin density (in units of $\hbar$) and $K$ is the anisotropy constant (in units of energy), which is easy plane when $K>0$ and easy axis when $K<0$. The spin density operator $\hat{\mathbf{s}}$ consists of a coherent piece $\langle\hat{\mathbf{s}} \rangle=\tilde{s} \mathbf{n}$ around which the spin density fluctuates incoherently: $\delta \hat{\mathbf{s}}=\hat{\mathbf{s}}-\langle \hat{\mathbf{s}}\rangle$. These fluctuations are composed of magnons, which reduce the effective spin density to $\tilde{s}=s(1-n_x/s)$, where $n_x$ is the thermal magnon density.  The anisotropy couples the thermal cloud and $\mathbf{n}$, allowing for an exchange of angular momentum between the two via the scattering of thermal magnons. The local mechanism for the thermally-driven torque works as follows:  When a temperature gradient is applied across the N$|$F interface, angular momentum is driven into the normal cloud by the spin Seebeck effect.  The out-of-equilibrium cloud relaxes the excess angular momentum to the order parameter via this coupling, thereby exerting a torque on $\mathbf{n}$.  

\subsubsection{Scattering} 
Using the Holstein-Primakoff transformation, the spin density may be mapped to boson field operators $\hat{\Phi}(\mathbf{x})$ and  $\hat{\Phi}^\dagger(\mathbf{x})$: \cite{Holstein:1940ta}
\begin{equation}
\hat{s}_{z}=\hat{\Phi}^{\dagger}\hat{\Phi}-s\,,\,\,\,\hat{s}_-(\mathbf{x})=\sqrt{2s-\hat{\Phi}^{\dagger}\hat{\Phi}}\hat{\Phi}\,,
\label{HP}
\end{equation}
where $\hat{s}_\pm=\hat{s}_x\pm i \hat{s}_y$ and $[\hat{\Phi}(\mathbf{x}),\hat{\Phi}^{\dagger}(\mathbf{x}')]=\delta (\mathbf{x}-\mathbf{x}')$. It is convenient to decompose $\hat{\Phi}$ into $\Phi=\langle \hat{\Phi}\rangle$, corresponding to a coherent condensate magnon, and $\hat{\varphi}$, which describes fluctuations around $\Phi$:
\begin{equation}
\hat{\Phi}(\mathbf{x})=\Phi+\hat{\varphi}(\mathbf{x})\, .
\label{coherent}
\end{equation} 
The quanta of $\hat{\varphi}$ are incoherent magnons, each of which carry angular momentum $-\hbar \mathbf{n}\approx \hbar \mathbf{z}$; for our purposes, these magnons are thermally activated, so that the thermal magnon cloud density is given by $n_x=\langle \hat{\varphi}^\dagger \hat{\varphi} \rangle$.  Writing $\Phi \equiv\sqrt{n_c} e^{-i \phi}$, which plays the role of the condensate wave function, the total average angular momentum in the $z$ direction becomes:  $\hbar \langle \hat{s}_z\rangle=\hbar(n_x+n_c)-\hbar s$. Using Eq.~(\ref{HP}) to compute the remaining components of $\langle\hat{\mathbf{s}} \rangle=\tilde{s} \mathbf{n}$, one identifies $\phi$ as the azimuthal angle between $\mathbf{n}$ and the $x$ axis;  we assume that $n_c+n_x\ll s$, so that the condensate density $n_c$, which parametrizes the misalignment of $\mathbf{n}$ with $-\mathbf{z}$, can be written as $n_c \approx (\tilde{s}/2)\theta^2 $, with $\theta$ as the polar angle between them (see Fig.~\ref{sch1}).

Expanding Eq.~(\ref{HP}) in $\hat{\Phi}^\dagger \hat{\Phi}/s$ and inserting Eq.~(\ref{coherent}) generates terms of various powers of $\hat{\varphi}$ and $\Phi$, which divide into two classes: $\hat{\varphi}$ magnon number conserving, and nonconserving.  In the absence of driving, the former terms relax the thermal magnon distribution $f_{\mathbf{k}}\equiv \langle \hat{\varphi}^\dagger_{\mathbf{k}} \hat{\varphi}_{\mathbf{k}} \rangle$ (with $\hat{\varphi}_{\mathbf{k}} =\int d^3 x e^{i\mathbf{k}\cdot \mathbf{x}} \hat{\varphi}/\sqrt{V}$ and $V$ as the volume of F) towards a Bose-Einstein profile: $f_{\rm{BE}}(\epsilon_{\mathbf{k}})=1/[e^{\beta(\epsilon_{\mathbf{k}}-\mu)}+1]$ with a well-defined magnon temperature $T=1/\beta$ (in units of energy) and chemical potential $\mu$. The magnon spectrum $\epsilon_{\mathbf{k}}=A\mathbf{k}^2+U$ is shifted by the Hartree-Fock mean-field potential $U=\hbar \Omega+2 Kn_c/s$, where $\hbar \Omega=H-K(1-2n_x/s)$. The relaxation time associated with these processes depends on both the exchange and anisotropy terms in Eq.~(\ref{H}).  Focusing on high temperatures ($T\gg \hbar \Omega$), the exchange mechanism dominates, and the relaxation time is fast;\cite{Bender:2014hga} we shall therefore suppose that the thermal magnon cloud is parametrized by $T$ and $\mu$, even in the presence of driving. In addition, in equilibrium Gilbert damping establishes $\mu=0$, while inelastic spin-preserving magnon-phonon scattering fixes the magnon temperature to that of the phonons.  Because the former type of magnon-lattice interaction relies on spin-orbit coupling, it is generally weaker than the latter; in this spirit, we shall suppose that the magnon temperature always remains pinned to that of the phonon temperature, while $\mu$ may be driven from its equilibrium value.

The spin torque on $\mathbf{n}$ arises from terms in $\hat{\mathcal{H}}$ that break $\hat{\varphi}$ magnon number conservation. The exchange interaction, which is independent of $\mathbf{n}$ by SU(2) symmetry, does not contribute.  However, when $\mathbf{n}$ is misaligned with the $z$ axis, anisotropy generates a contribution to $\hat{\mathcal{H}}$,
\begin{equation}
\hat{\mathcal{H}}_{\mathrm{cx}}({\mathbf{n}})=(K/s)\Phi^* \int d^3 x \hat{\varphi}^\dagger(\mathbf{x}) \hat{\varphi}(\mathbf{x}) \hat{\varphi}(\mathbf{x})+H.c. \, ,
\label{scham}
\end{equation} 
opening a magnon scattering channel that redistributes $z$ angular momentum between the thermal cloud and order parameter: two $\hat{\varphi}$ magnons are annihilated (created), creating (annihilating) one $\hat{\phi}$ magnon and one $\Phi$ magnon. Because the total $z$-angular momentum carried by the spin density is conserved by rotational symmetry, the corresponding loss (gain) of $\hbar \mathbf{z}$ angular momentum by the thermal magnon cloud is compensated by the absorption (emission) of angular momentum by $\Phi$, which is translated as an antidamping (damping) spin torque on $\mathbf{n}$. The resulting scattering rate at which angular momentum is transferred between the thermal cloud and the order parameter is obtained by Fermi's Golden Rule and given by:\cite{Zaremba:1999iu}
\begin{equation}
\Gamma=2 \alpha_{\rm{cx}}\left( \hbar \omega-\mu\right) n_c=\left.\hbar \dot{n}_x \right|_{\mathrm{cx}}=-\left.\hbar \dot{n}_c \right|_{\mathrm{cx}}\, .
\label{nrate}
\end{equation} 
Here, $f_i=f_{\mathrm{BE}}(\epsilon_{\mathbf{k}_i})$, $\hbar \omega \equiv \hbar \Omega +K n_c/s$ is the precessional frequency of $\mathbf{n}$ around $-\mathbf{\mathbf{z}}$, and $\alpha_{\rm{cx}}$ is given by:
\begin{align}
\alpha_{\rm{cx}}&=\frac{(K/s)^{2}}{T(2\pi)^{5}}\int d^{3}k_{1}\int d^{3}k_{2}\int d^{3}k_{3}\delta\left(\mathbf{k}_{1}-\mathbf{k}_{2}-\mathbf{k}_{3}\right)\label{alphasc} \\
&\times\delta\left(\hbar \omega+\epsilon_{\mathbf{k}_1}-\epsilon_{\mathbf{k}_2}-\epsilon_{\mathbf{k}_3}\right)\left(1+f_{1}\right)f_{2}f_{3}\,. \nonumber
\end{align}
which can be written as: $\alpha_{\rm{cx}}=\bar{\alpha}_{\rm{cx}}I$, where $\bar{\alpha}_{\rm{cx}}\equiv (T/T_c)^3(K/T)^2$, $T_c=As^{2/3}$ approximates the Curie temperature, and $I$ is a dimensionless integral that depends on $(\mu-U)/T$ and $\hbar \omega/T$.  According to Eq.~(\ref{nrate}), when $\mu=\hbar \omega$, the thermal cloud and order parameter are in equilibrium, corresponding to an entropic maximum of the closed magnetic subsystem of F.

\subsubsection{Driven magnetic dynamics}

At zero temperature, the thermal cloud is absent, and the dynamics of the order parameter of F are described by the Landau-Lifshitz-Gilbert phenomenology:
\begin{equation}
(1+\alpha \mathbf{n}\times)\hbar \dot{\mathbf{n}}+\mathbf{n}\times \mathbf{H}= (\alpha'_i+\alpha'_r \mathbf{n}\times)( \boldsymbol{\mu}'\times\mathbf{n}- \hbar\dot{\mathbf{n}})\, ,
\label{LLG}
\end{equation}
where $\mathbf{H}=(H +K\mathbf{z}\cdot \mathbf{n} ) \mathbf{z}$ is comprised of the applied field $H$ and the anisotropy field $K \mathbf{z}\cdot \mathbf{n}$, and $\alpha$ is the bulk Gilbert damping of F, describing the flow of angular momentum from $\mathbf{n}$ to the lattice.  The quantities $\alpha'_r$ and $\alpha'_i$ describe angular momentum transfer with the normal metal N and are the real and imaginary parts of $g^{\uparrow \downarrow}/4\pi s d_F$, where $g^{\uparrow \downarrow}$ is the spin-mixing conductance at the interface and $d_F$ is the thickness of F. In general, the spin accumulation $\boldsymbol{\mu}'$ must be self-consistently determined by complementing magnetic dynamics with a treatment of spin transport in N; we shall circumvent this inessential complication by taking N to be a good spin sink, so that the spin accumulation is electrically tunable (by e.g. the spin Hall effect) independently of the temperature gradient\cite{Note1} and chosen to be along the $z$-axis: $\boldsymbol{\mu}'=\mu'\mathbf{z}$. Provided $H-K>0$, the equilibrium ($\mu'=0$) solution to Eq.~(\ref{LLG}) is $\mathbf{n}=-\mathbf{z}$.

At finite temperatures, the scattering by the thermal cloud of magnons modifies the order parameter dynamics, and the coefficients $\alpha'_r$ and $\alpha'_i$ acquire temperature dependent corrections $\sim n_x/s$.  The temperature dependent magnon gap $\hbar \Omega$ must be positive (which we will assume through the remainder of Sec.~\ref{secloc}) for $\mathbf{n}=-\mathbf{z}$ ($n_c=0$) to be a stable equilibrium in the absence of driving.   Here, it is convenient to recast Eq.~(\ref{LLG}) as a rate equation for $n_c$, to which the scattering rate $\Gamma$ is phenomenologically added; denoting $\alpha_r'=\alpha'$, and neglecting higher order terms in the $\alpha$'s, one has, for small angle dynamics ($\theta \ll 1$):
\begin{equation}
\hbar \dot{n}_c=2\alpha ' \mu'n_c-2(\alpha+\alpha')\hbar \omega n_c-\Gamma \, ,
\label{ncdot}
\end{equation}
where the first two terms on the right-hand side follow from Eq.~(\ref{LLG}), while $\Gamma$ is given by Eq.~(\ref{nrate}).   Eq.~(\ref{ncdot}) can be recast back as a finite-temperature Landau-Lifshitz-Gilbert equation (valid for small angle dynamics):
\begin{equation}
(1+\alpha \mathbf{n}\times)\hbar \dot{\mathbf{n}}+\mathbf{n}\times \tilde{\mathbf{H}}= \alpha' \mathbf{n}\times( \boldsymbol{\mu}'\times\mathbf{n}- \hbar\dot{\mathbf{n}})+\boldsymbol{\tau}_l\, ,
\label{LLGT}
\end{equation}
which is one of the central results of this paper.  Here, $\boldsymbol{\tau}_l=\alpha_{\rm{cx}} \mathbf{n}\times(\boldsymbol{\mu}\times \mathbf{n}-\hbar \dot{\mathbf{n}})$ is the local spin torque, with $\boldsymbol{\mu}=\mu\mathbf{z}$, and $\tilde{\mathbf{H}}=[\hbar \Omega+K (1+\mathbf{z}\cdot\mathbf{n})]\mathbf{z}$.

Complementing the dynamics of the order parameter, Eq.~(\ref{ncdot}), is that of thermal cloud. In response to a spin accumulation $\boldsymbol{\mu}'$ and/or a temperature gradient, angular momentum in the $-\mathbf{n}$ direction is driven into (out of) F and absorbed (emitted) by the thermal cloud, creating an out-of-equilibrium chemical potential $ \mu>0$ ($<0$). For fixed magnon temperature, the rate of change of the chemical potential resulting from these biases may be obtained from the total rate equation for the thermal-cloud density:
\begin{equation}
\hbar \dot{n}_x=\dot{\mu}\partial_\mu n_x =j_\parallel/d_F-g \mu/d_F+\Gamma \, .
\label{ndot}
\end{equation}
In the first term on the right-hand side, $j_\parallel$ is the current injected from N across the interface, which in linear response is given by:
\begin{equation}
j_\parallel=g'  (\mu'-\mu)+\mathcal{S}'\delta T \, .
\label{intcurrent}
\end{equation}
The quantities\cite{Bender:2015jx} $g'\sim g^{\uparrow \downarrow} (T/T_c)^{3/2}$ and $\mathcal{S}'\sim g'$ are the temperature-dependent interfacial magnon conductance and spin Seebeck coefficients, which are both proportional to $g^{\uparrow\downarrow}$; $\delta T=T'-T$ is the difference between the electron temperature $T'$ and magnon temperature $T$. The second term on the right-hand side of Eq.~(\ref{ndot}), parametrized by $g\sim \alpha$, describes the Gilbert damping of thermal-cloud angular momentum into the F lattice, which, in the absence of driving by N relaxes $\mu$ to zero. 

Together, Eqs.~(\ref{ncdot}) and~(\ref{ndot}) form a closed set of coupled equations for the condensate density $n_c$ and the thermal cloud chemical potential $\mu$.  Separating the ``fast" dynamics of the thermal magnons from the ``slow" dynamics of the order parameter, we solve Eq.~(\ref{ndot}) for the magnon steady-state condition $\dot{n}_x=0$ to obtain a chemical potential $\mu=(\mathcal{S}' \delta T+g'\mu'+2\alpha_{cx}n_c \hbar \omega d_F)/(g+g'+2 d_F \alpha n_c)$.

\subsubsection{Ferromagnetic resonance linewidth}
\label{LR}

Focusing on behavior near equilibrium ($\mu'=\delta T=\mu=n_c=0$), Eq.~(\ref{ncdot}) may be written as $\hbar\dot{n}_c=2\alpha_{\rm{tot}} \hbar \omega n_c$, where $\alpha_{\rm{tot}}=\alpha+\alpha'+\alpha_{\rm{cx}}-(\alpha' \mu'+\alpha_{\rm{cx}}\mu)/\hbar \omega$ is the total damping of $\mathbf{n}$, and $\alpha_{cx}$ is evaluated in equilibrium ($\mu=n_c=0$). To lowest order in $n_c$, $\mu=(\mathcal{S}'\delta T+g' \mu')/(g+g')$, which when inserted into Eq.~(\ref{ncdot}) yields $\alpha_{\rm{tot}}=\alpha+\alpha'+\alpha_{\rm{cx}}+\Delta \alpha_{\mu'}+\Delta \alpha_{T}$, governing the relaxation of $n_c$. The contribution $\Delta \alpha_{\mu'}=-\left[\alpha'+\alpha_{\rm{cx}}/(1+g/g')\right](\mu'/\hbar \Omega)$ consists of the zero-temperature term $\propto \alpha'$ and the thermal enhancement $\propto \alpha_{\rm{cx}}$. The change in damping resulting from a temperature gradient,
\begin{equation}
\Delta \alpha_{T}=-\alpha_{\rm{cx}} \frac{\mathcal{S}'}{g+g'} \frac{\delta T}{\hbar \Omega}\, ,
\label{deltaalphaT}
\end{equation}
is due to the thermal magnons in its entirety and is one of the central results of this paper. Both $\Delta \alpha_{\mu'}$ and $\Delta \alpha_{T}$ can be deduced from ferromagnetic resonance measurements. 

\subsubsection{DC-pumped magnon condensates}
\begin{figure}[pt]
\includegraphics[width=0.85\linewidth,clip=]{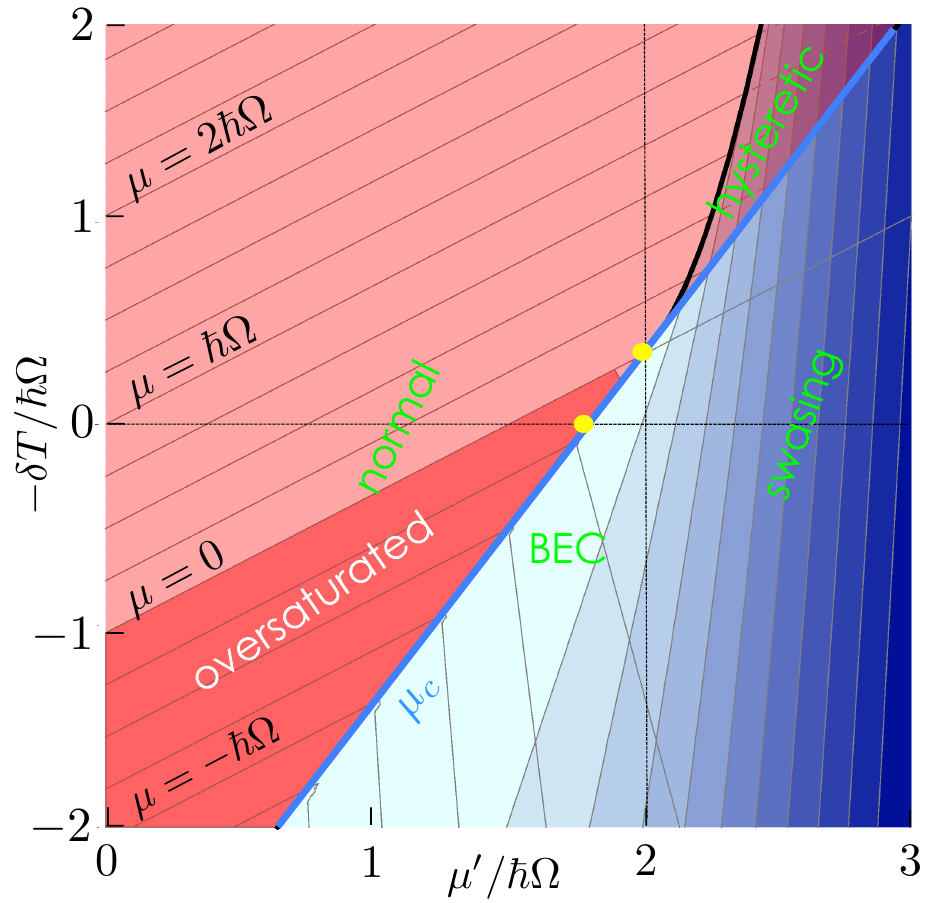}
\caption{Phase diagram for F$|$N bilayer corresponding to the stable solutions to the coupled Eqs.~(\ref{ncdot}) and~(\ref{ndot}), demonstrating normal phase, Bose-Einstein condensation (BEC) and swasing ($\mu'>\hbar \Omega[1+(\alpha+\alpha_{\mathrm{sc}})/\alpha'$]).  Here we have taken $\alpha_{\mathrm{sc}}=\alpha/2=\alpha'/2$, $K=\hbar \Omega$, $T=10^2 \hbar \Omega$ and $s (A/\hbar \Omega)^{3/2}=10^4$. When $\mu> \hbar \Omega$ (below the phase transition), the thermal cloud is oversaturated. (See Ref.~\onlinecite{Bender:2014hga}).}
\label{AlphascPDfig}
\end{figure}

Finite-angle dynamics of $\mathbf{n}$ may be excited upon the application of a sufficiently large spin accumulation and/or temperature gradient.  This was the subject of Ref.~\onlinecite{Bender:2014hga}, in the limit in which the condensate and cloud are strongly coupled ($\alpha_{\rm{cx}}\rightarrow \infty$).  There, when F is in normal phase ($n_c=0$), $\mu<\hbar \Omega$ is determined from the steady state condition $ \dot{n}_x=0$, as in Sec.~\ref{LR}; when F is in condensate phase ($n_c>0$), the Bose-Einstein gas of thermal magnons becomes saturated ($\mu=\hbar \Omega$), and $n_c$ is determined from the steady state condition $\dot{n}_c=0$. Together, Eqs.~(\ref{ncdot}) and~(\ref{ndot}) represent a generalization of Ref.~\onlinecite{Bender:2014hga} to finite cloud-condensate coupling, with the structure of the phase diagram determined by the steady-state solutions for $\mu$ and $n_c$ to the joint condition $\dot{n}_c=\dot{n}_x=0$.  In the strong condensate-cloud coupling regime ($\alpha_{\rm{cx}}\gg \alpha, \alpha'$), the phase diagram of Ref.~\onlinecite{Bender:2014hga} is reproduced.  More generally, the cloud chemical potential must overcome a threshold $\mu_c =\left[1+(\alpha+\alpha')/\alpha_{\rm{cx}} \right] \hbar \Omega-\alpha'\mu'/\alpha_{\rm{cx}}$ in order to realize a steady-state condensate; thus, the cloud may become oversaturated, with $\mu>\hbar \Omega$,\cite{Note2} but damping by the lattice and N relaxes angular momentum of the condensate more quickly than it is replenished by cloud-condensate scattering when $\mu<\mu_c$, so that F remains in normal phase.  The corresponding phase diagram is shown in Fig.~\ref{AlphascPDfig}, with the terminology borrowed from Ref.~\onlinecite{Bender:2014hga}.

\section{Nonlocal mechanism}
\label{secnl}
\begin{figure}[pt]
\includegraphics[width=\linewidth,clip=]{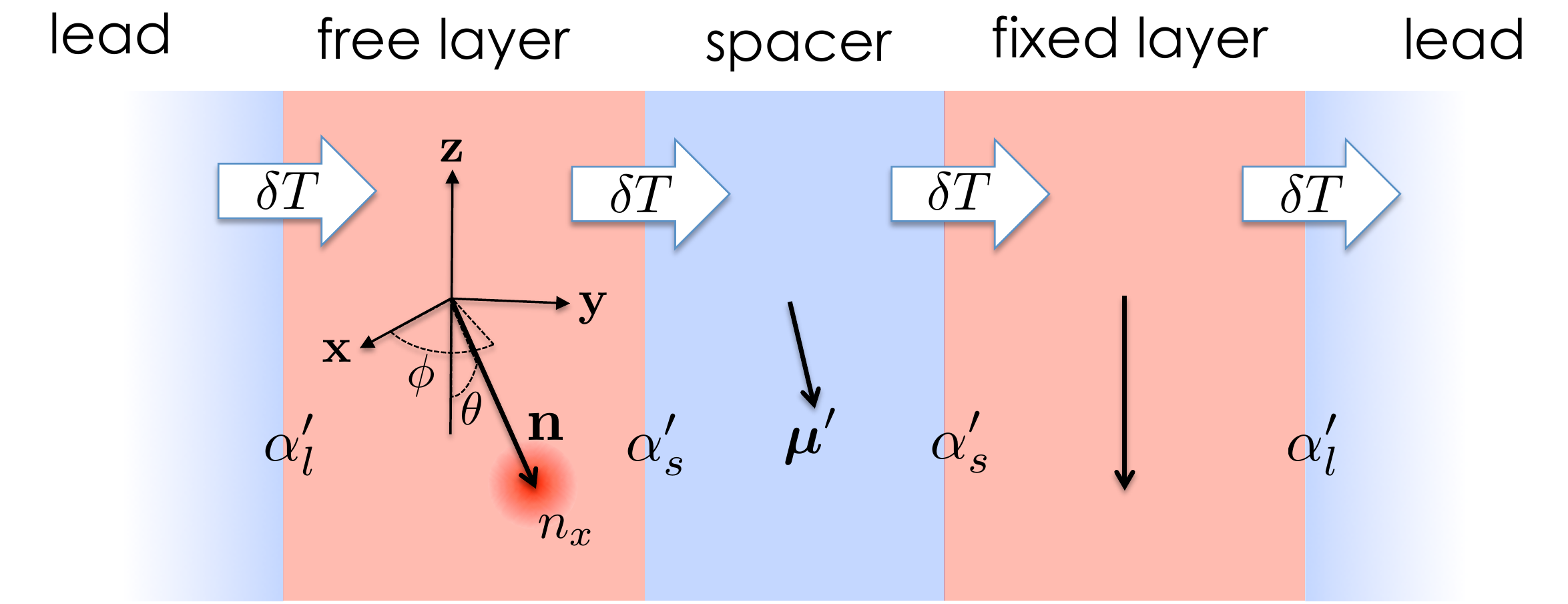}
\caption{Thermally-biased spin valve.  A heat flux drives spin accumulation $\boldsymbol{\mu}'$ (in the plane defined by $\mathbf{n}$ and $\mathbf{z}$) into the normal metal spacer. When free layer spin density is misaligned with the $\mathbf{z}$ axis, $\boldsymbol{\mu}'$ is no longer collinear with $\mathbf{n}$, and the component of  $\boldsymbol{\mu}'$ perpendicular to $\mathbf{n}$ provides a torque.}
\label{sch2}
\end{figure}

The second mechanism for thermal spin-transfer torque relies on the presence of an additional ferromagnetic layer to provide the SU(2) symmetry breaking required to realize a torque on $\mathbf{n}$. Let us now consider the simplest such structure: a spin valve, composed of two ferromagnet layers (one free and one fixed) separated by a normal metal spacer, as depicted in Fig.~\ref{sch2}.   In a conducting spin valve, electrical or thermal biasing generates a two channel spin current,\cite{Johnson:1985do,*Uchida:2010hy,Luc:2014hr} carried by electrons parallel and antiparallel to the order parameter in the magnetic layers, which exerts a torque on the free layer. 

In our electrically insulating structure, thermal biasing (applied perpendicularly to plane) generates a pure spin current that is single channel, carried through the ferromagnetic layers by magnons; this current results in a nonlocal torque on the free magnetic layer order parameter $\mathbf{n}$ as a consequence of the misalignment of the free and fixed layers.  Slonczewski\cite{Slonczewski:2010jh} has proposed a similar scheme. There, a heat current is converted into a spin current via a ferrite layer, which is coupled to a paramagnetic monolayer by superexchange; spin current is subsequently transferred to the conduction electrons of a spacer and ultimately to a free magnet. In contrast, our proposal relates the thermal spin-flux directly to the spin-mixing conductance, a readily measurable quantity, circumventing the need for a paramagnetic monolayer.

The spin valve we consider is a five layer structure (N-lead$|$free-F$|$N-spacer$|$fixed-F$|$N-lead), the mirror symmetry of which is broken by the fact that the pinned layer is fixed (e.g. by exchange biasing) with spin density oriented in the $-\mathbf{z}$ direction.  For simplicity, we assume all of the transport coefficients for the free and fixed layers to be identical. In equilibrium, the free layer is oriented either parallel or antiparallel to fixed layer.  In order to maximize the efficiency of spin transport across the structure, let us assume that the thickness  $d_F$ of each of the (monodomain) ferromagnet layers is much shorter than the thermal magnon diffusion length.  Likewise, we take the normal-metal spacer thickness $d_s$ to be much shorter than the electronic spin diffusion length therein, which may be accomplished by using a poor spin sink (e.g. Cu).  In contrast, let us for simplicity assume that the normal-metal leads attached to the ferromagnets are excellent spin sinks, such as Pt, so that no spin accumulates inside them. Spin transport between the magnetic layers and the spacer ($s$) is parametrized by the spin-mixing conductance $g^{\uparrow \downarrow}_s$, and between F and the leads ($l$) by $g^{\uparrow \downarrow}_l$.

In response to a temperature gradient applied across the structure, a spin current of magnons flows through the magnetic layers, resulting in a spin accumulation $\boldsymbol{\mu}'$ in the normal metal spacer. By symmetry, when the magnetic moments of the two ferromagnetic layers are parallel, $\boldsymbol{\mu}'$ vanishes; for all other orientations, $\boldsymbol{\mu}'$ builds up in the plane defined by $\mathbf{n}$ (the free layer spin density order parameter) and $\mathbf{z}$ (the fixed layer). When the two magnetic layers are misaligned, the spin accumulation exerts a damping-like nonlocal torque $\boldsymbol{\tau}_{nl}=- \alpha_s' \mathbf{n}\times  \mathbf{n}\times \boldsymbol{\mu}'$ on the free layer (with $\alpha_s'=\Re g_s^{\uparrow \downarrow}/4\pi s d_F$ as the effective Gilbert damping coefficient due to contact with the spacer).

Following the approach of Sec.~\ref{secloc} of separating order parameter and magnonic timescales and focusing on the latter, we have the magnon spin current density $j_i$ into ferromagnetic layer $\mathrm{F}_i$, for a fixed orientation of $\mathbf{n}$, is (with $i=1$ as the free layer and $i=2$ the fixed layer):
\begin{equation}
\hbar \dot{n}_i d_F=j_i=j_{l\rightarrow i}+j_{s\rightarrow i}+\tilde{j}_i\, ,
\label{ji}
\end{equation}
where $n_i$ is the thermal cloud magnon density in layer $i$, $j_{l\rightarrow i}=-g_l' \mu_i-(-1)^{i}\mathcal{S}_l' \delta T$ is the current entering F$_i$ from the lead, $j_{s\rightarrow i}=-g_s' (\mu_i+\mathbf{n}_i\cdot \boldsymbol{\mu})+(-1)^{i}\mathcal{S}_s' \delta T$ is the spin current entering $\mathrm{F}_i$ from the spacer, $\tilde{j}_i=-g \mu_i$ the spin current lost to Gilbert damping of thermal magnons, $\mu_i$ is the thermal magnon chemical potential in each ferromagnet, $\mathbf{n}_1=\mathbf{n}$ and $\mathbf{n}_2=-\mathbf{z}$.  The quantities $g_l '$ and $g_s '$ are the magnon conductances of the interfaces of F$_i$ with the leads ($l$) and spacer ($s$), respectively, and $\mathcal{S}_l'$ and $\mathcal{S}_s '$ are the spin Seebeck coefficients of the interfaces of F with the leads and spacer. While a thorough treatment of spin transport involves a detailed account of how magnon, phonon and electron temperature profiles are established throughout the structure, we have assumed, for our proof-of-principle calculation, that the electron/magnon temperature difference $\delta T$ is the same across all interfaces (see Fig.~\ref{sch2}).  The rate of change of the spin density $\boldsymbol{\rho}=D_F \boldsymbol{\mu}(\hbar/2)$ (with $D_F$ as the Fermi surface density of states) inside the spacer is:
\begin{equation}
\dot{\boldsymbol{\rho}} d_s=(-\mathbf{n})j_{1 \rightarrow s}+\mathbf{z} j_{2 \rightarrow s}\, ,
\label{js}
\end{equation}
with $j_{i\rightarrow s}=-j_{s\rightarrow i}$. In a steady state of magnon flux, we require $\dot{n}_1$, $\dot{n}_2$ and $\dot{\boldsymbol{\rho}}$ to vanish, which, employing Eqs.~(\ref{ji}) and~(\ref{js}), yields a closed set of five equations for $\mu_1$, $\mu_2$ and $\boldsymbol{\mu}'$. Solving for the latter and inserting into $\boldsymbol{\tau}_{nl}=- \alpha_s' \mathbf{n}\times  \mathbf{n}\times \boldsymbol{\mu}'$ yields the thermally-driven torque on $\mathbf{n}$.

 In order to characterize linear response and magnetic dynamics, it suffices to find $\boldsymbol{\tau}_{nl}$ near the parallel configuration of the free and fixed layers ($\mathbf{n} = -\mathbf{z}$) and the antiparallel configuration ($\mathbf{n} =\mathbf{z}$). Near the parallel configuration, we obtain
\begin{equation}
\boldsymbol{\tau}_{nl}\approx  \sigma _{p} \delta T \mathbf{n}\times\mathbf{n}\times\mathbf{z}\, ,
\label{ptau}
\end{equation}
where
\begin{align}
\sigma_{p} =\alpha_s '\frac{(g+g_l')\mathcal{S}_s'+g_s' \mathcal{S}_l'}{2 g_s'(g+g_s'+g_l')} \, ;
\label{alphap}
\end{align}
near the antiparallel configuration,
\begin{equation}
\boldsymbol{\tau}_{nl}\approx  \sigma _{ap} \delta T \mathbf{n}\times\mathbf{n}\times \mathbf{z}
\label{aptau}
\end{equation}
where 
\begin{equation}
\sigma_{ap}=\frac{\alpha_s '}{2}\left(\frac{\mathcal{S}_s'}{g_s'}+\frac{\mathcal{S}_l'}{g+g_l' }\right)\, .
\label{alphaap}
\end{equation} 

The dynamics of the free layer are captured by the LLG equation:
\begin{equation}
(1+\alpha \mathbf{n}\times)\hbar \dot{\mathbf{n}}+\mathbf{n}\times \mathbf{H}=-\hbar (\alpha'_{l}+\alpha'_{s} )\mathbf{n}\times \dot{\mathbf{n}}+\boldsymbol{\tau}_{nl}\, ,
\label{llg2}
\end{equation}
where we've assumed $\boldsymbol{\tau}_{l}\ll \boldsymbol{\tau}_{nl}$ so that the local torque is neglected. In order to characterize the small angle dynamics of the free layer near the poles $\mathbf{n}=\pm \mathbf{z}$, we parametrize $\mathbf{n}$ by spherical coordinates as above and expand Eq.~(\ref{llg2}) in $\theta$, neglecting for simplicity the local torque $\boldsymbol{\tau}_{l}$. Near the parallel configuration ($\theta=0$), we obtain an equation of motion $\hbar \dot{\theta}\approx-\epsilon_p \theta$, where $\epsilon_p=\alpha_p \hbar \Omega$, with $\alpha_p=\alpha'_l+\alpha'_s+\alpha+\Delta \alpha_p$ as the total damping and
\begin{equation}
\Delta \alpha_{p}=\sigma_p \delta T/\hbar \Omega\, ,
\label{Dalphap}
\end{equation}
as the change in damping near the parallel configuration resulting from the $\delta T$. Near the antiparallel configuration ($\theta=\pi$), $\hbar \dot{\theta}\approx \epsilon_{ap} (\pi-\theta)$, where $\epsilon_{ap}=\alpha_{ap}\hbar \tilde{\Omega}$, with $\hbar \tilde{\Omega}=H+K(1-2n_x/s)$ as the magnon gap there, $\alpha_{ap}=\alpha'_l+\alpha'_s+\alpha+\Delta \alpha_{ap}$ as the total damping, and 
\begin{equation}
\Delta \alpha_{ap}=- \sigma_{ap} \delta T/\hbar \tilde{\Omega}
\label{Dalphaap}
\end{equation}
as the change in damping near the antiparallel configuration. 

\begin{figure}[pt]
\includegraphics[width=\linewidth,clip=]{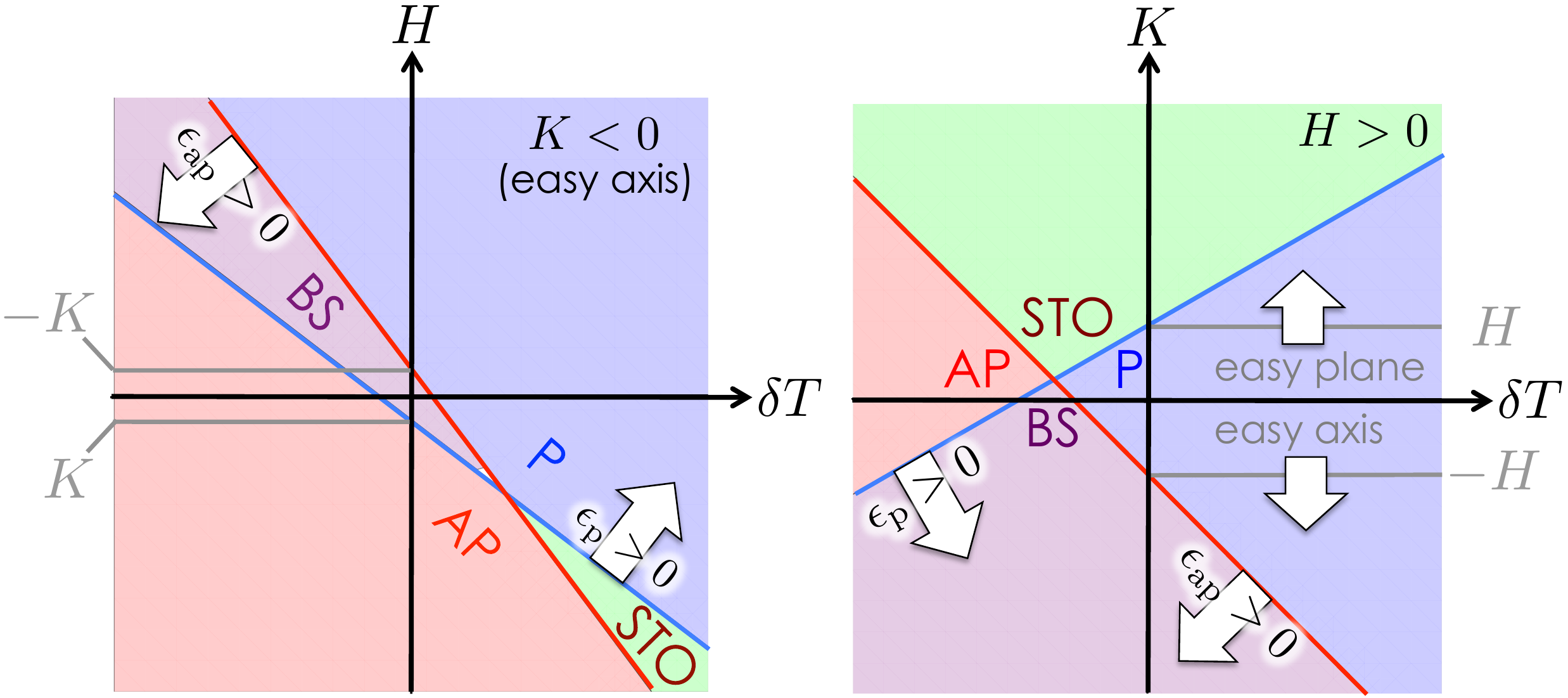}
\caption{Phase diagrams for the free layer in the spin valve for constant $K(<0)$ and constant $H(>0)$, showing parallel ($\mathrm{P}$),  antiparallel (AP), bistable (BS) and spin-torque oscillator (STO) phases.} 
\label{svpp}
\end{figure}

Beyond linear response, the spin valve can be driven into different phases, the boundaries of which are defined by the planes $\epsilon_{p}=0$ and $\epsilon_{ap}=0$ in $H-K-\delta T$ space. When both $\epsilon_{\mathrm{p}}$ and $\epsilon_{\mathrm{ap}}$ are positive, the free layer is bistable. When $\epsilon_{\mathrm{p}}$ is positive (negative) and $\epsilon_{\mathrm{ap}}$ is negative (positive), the free layer is stabilized in the $-\mathbf{z}$ ($+\mathbf{z}$ direction).  Last, when both $\epsilon_{\mathrm{p}}$ and $\epsilon_{\mathrm{ap}}$ are negative, the dynamics stabilize to a limit cycle with $0<\theta<\pi$, i.e. the magnet is a spin-torque oscillator.   The resulting phase diagram is show in Fig.~\ref{svpp}.

\section{Conclusion}

We have shown how both local and nonlocal thermally-driven spin torques $\boldsymbol{\tau}_l$ and $\boldsymbol{\tau}_{nl}$ may arise in magnet/metal heterostructures, which modify the magnetic dynamics [Eqs.~(\ref{LLGT}) and~(\ref{llg2})].   In linear response, these torques manifest in both the bilayer and spin valve as changes in the damping [Eqs.~(\ref{deltaalphaT}),~(\ref{Dalphap}) and~(\ref{Dalphaap})]; at a threshold bias for each structure, the net effective damping reverses its sign, resulting in finite-angle dynamics. For the case of the spin valve, the change in the damping of the free layer, near either orientation, goes as $\Delta \alpha \sim \alpha' _s \delta T/\hbar \Omega$.  When F is sufficiently thin, $\alpha_s'$ becomes comparable to $\alpha$, so that a temperature difference $\delta T \sim \hbar \Omega$ results in a change in damping $\Delta \alpha$ that can overcome the intrinsic Gilbert damping; taking $\alpha \sim 10^{-4}$ and using yttrium-iron-garnet (YIG) magnetic layers with platinum leads, this is the case when the thickness of F is less than $\sim$100 nm.  For an F$|$N bilayer, which relies on the local mechanism, assuming again that $\alpha'>\alpha$, the change in damping goes as $\sim \alpha_{\rm{cx}} \delta T/\hbar \Omega$. Supposing that $K\approx 4\pi M_s^2/s$ corresponds to shape anisotropy, and using YIG parameters (with $M_s\approx 150 $ emu/cm$^3$ as the saturation magnetization), one arrives at $\bar{\alpha}_{\rm{cx}}\approx 10^{-6}$, which may be further enhanced by the factor $I$. 

Our model makes several assumptions. First, we rely on strong magnon-magnon scattering to thermalize the cloud of incoherent magnons to a Bose-Einstein profile. At low temperatures and high driving, the magnons may no longer be parametrizable by a local chemical potential and temperature.  Additionally, implicit in our treatment is the assumption of strong inelastic spin-preserving magnon-phonon coupling that fixes the magnon temperature to that of the phonons, which is taken to increase in a steplike fashion across the structure.  If the magnon-phonon coupling is not sufficiently strong, the magnon temperature profile must be determined from an appropriate theory for the magnon heat transport. Likewise, in order to make quantitative experimental predictions, a more detailed account of how heating in the normal metals establishes a phononic temperature profile across the structure is necessary. We have exploited the separation of magnonic and order-parameter dynamical time scales, which may need to be examined more carefully in practice. Last, the scattering rate coefficient $\alpha_{\rm{cx}}$ depends sensitively on the low-energy magnon spectrum and inelastic scattering that may not be sufficiently strong to achieve full thermalization; further work is warranted to address the problem of the magnon equilibration at the bottom of the magnon band, for a given material under consideration. Future efforts may also apply and extend our approach to the problem of thermal spin torques to new structures, e.g., more complex multilayers, superlattices, and antiferromagnets. 

This work was supported by the US DOE-BES under Award No. DE-SC0012190.

%

\end{document}